\documentclass[useAMS,usenatbib,usegraphicx
]{mn2e}

\usepackage[setpagesize=false]{hyperref}
\providecommand{\adsurl}[1]{\href{#1}{ADS}}

\usepackage{braket,amsmath}
\input{colordvi.tex}

\title[Scale-dependent bias due to primordial vector fields]{Scale-dependent bias due to primordial vector fields}
\author[Maresuke Shiraishi, Shuichiro Yokoyama, Kiyotomo Ichiki and Takahiko Matsubara]{Maresuke Shiraishi$^{1}$\thanks{E-mail:
mare@nagoya-u.jp}, 
Shuichiro Yokoyama$^{2}$
, 
Kiyotomo Ichiki$^{1, 3}$
\newauthor
and Takahiko Matsubara$^{1, 3}$
\\
$^{1}$Department of Physics and Astrophysics, Nagoya University, Nagoya, Aichi, 464-8602, Japan\\
$^{2}$Institute for Cosmic Ray Research, The University of Tokyo, 5-1-5 Kashiwa-no-Ha, Kashiwa, Chiba, 277-8582, Japan\\
$^{3}$Kobayashi-Maskawa Institute for the Origin of Particles and the Universe, Nagoya University, Nagoya, Aichi, 464-8602, Japan
}
\begin{document}

\date{}

\pagerange{\pageref{firstpage}--\pageref{lastpage}} \pubyear{2012}

\maketitle

\label{firstpage}

\begin{abstract}
Anisotropic stress perturbations induced by primordial Gaussian vector fields create non-Gaussianity in curvature perturbations. 
We found that such non-Gaussianity closely resembles the local-type non-Gaussianity parametrized by $f_{\rm NL}$, and generates scale-dependent bias in large-scale structures. 
We also found a simple relationship between the scale-dependent bias and the power spectrum of the vector fields. When the vector fields are interpreted as primordial magnetic fields, the effective $f_{\rm NL}$ is shown to be always negative. The scale-dependent bias provides a new approach to probing primordial vector fields. 
\end{abstract}


\section{Introduction}

Cosmological models with vector fields have often been considered in order to explain the origin of large-scale cosmic magnetic fields and the statistical anisotropy of the Universe (e.g. \citet{Widrow:2002ud, Bamba:2006ga, Martin:2007ue, Watanabe:2010fh}), although a successful model of inflationary magnetogenesis has yet to be derived, as shown by \citet{Bamba:2003av, Kanno:2009ei, Demozzi:2009fu, Demozzi:2012wh, Suyama:2012wh, Fujita:2012rb}. Recently, beyond the framework of cosmological magnetogenesis, global analyses of such cosmological vector fields involving higher-order cosmological perturbations have attracted attention, as they provide interesting imprints (e.g. \citet{Ackerman:2007nb, Watanabe:2010bu, Sorbo:2011rz, Barnaby:2012xt}).


The non-Gaussianity of primordial fluctuations has recently been considered as a powerful observable with which to extract reliable results from the huge number of models of the early Universe (for a review, see for example \citet{Bartolo:2004if, Komatsu:2010hc}). One can see beneficial bounds from the higher-order correlations of cosmic microwave background (CMB) fluctuations on not only the amplitude but also the spectral tilt of the non-Gaussianity, which depend on theoretical parameters associated with scalar perturbations (e.g. \citet{Sefusatti:2009xu, Komatsu:2010fb, Smidt:2010ra, Hikage:2012bs, Becker:2012je, Becker:2012yr}). Many studies have also evaluated the non-Gaussian impacts of not only pure scalar perturbations but also vector perturbations driven by vector fields (e.g., \citet{Brown:2005kr, Yokoyama:2008xw, Karciauskas:2008bc, ValenzuelaToledo:2009nq, Dimastrogiovanni:2010sm, Barnaby:2010vf, Barnaby:2011vw, Barnaby:2012tk, Caldwell:2011ra, Motta:2012rn, Jain:2012ga, Bartolo:2012sd}). Such vector perturbations generate non-trivial non-Gaussianities and they have been investigated using the CMB power spectrum and bispectrum (\citet{Shiraishi:2012rm, Shiraishi:2012sn, Shiraishi:2012xt}). 


Together with the CMB spectra, the scale dependence of the halo/galaxy bias has been investigated as a good estimator of the primordial non-Gaussianity of curvature perturbations (see e.g. \citet{Slosar:2008hx, Verde:2010wp}). 
Recently, much effort has been put into directly exploring the scale dependence of the primordial bispectrum via the bias parameter. While the bias parameter generated from the exact local-type non-Gaussianity has a $k^{-2}$ dependence, there are models that predict the deviation from this dependence (e.g. \citet{Shandera:2010ei}). The shape of the bispectrum of curvature perturbations arising from the vector fields directly reflects the tilt of the power spectrum of the vector fields, which depends strongly on the generation mechanism of the vector fields. In this sense, the scale dependence of the bias parameter is expected to afford a useful clue to the nature of the primeval vector fields. 

This paper focuses on the impacts of the non-Gaussianity induced by the vector fields on the scale-dependent bias. To estimate this, we assume the existence of the Gaussian vector fields in the early Universe. Accordingly, the anisotropic stress perturbations arising from the square of these Gaussian fields obey chi-square statistics and create non-Gaussian curvature perturbations. In this paper, we adopt such vector-induced curvature perturbations as a source of the scale-dependent bias. Through the analysis of the bispectrum shape, we find that such a non-Gaussianity closely resembles a local-type configuration and hence we derive a relation between the amplitudes and spectral indices of the vector fields and the local-type nonlinearity parameter. By applying a general formalism for the scale-dependent bias induced by the primordial bispectrum, which is based on the integrated perturbation theory (iPT) (\citet{Matsubara:2011ck}) and discussed in \citet{Matsubara:2012nc}, we compute the bias parameter from the vector fields and confirm the validity of the above relation in the scale-dependent bias for several redshifts. At the same time, we observe tiny deviations from the local-type bias on small scales. Applying the formulae in the case where the vector fields are interpreted as magnetic fields, we obtain the interesting result that the scale-dependent bias has negative values. 

This paper is organized as follows. In the next section, we present a generation mechanism for the non-Gaussian curvature perturbations induced by the primordial vector fields. In Section~\ref{sec:bias}, we compute the scale-dependent bias originating from such a non-Gaussianity for several spectral indices of the vector fields and redshifts. The final section contains a summary and discussion.

\section{Primordial non-Gaussianity generated from vector fields}\label{sec:non-gaussianity}


In this section, we first discuss the non-Gaussian curvature perturbations, which are generated from the anisotropic stress perturbations induced by various types of vector fields in the early Universe. Next, we show that electromagnetic fields, which may occur in inflation, can produce such non-Gaussian anisotropic stress perturbations. 

\subsection{Curvature perturbations induced by anisotropic stress perturbations}

If the anisotropic stress perturbations, which scale like radiations ($\propto a^{-4}$ with $a$ being the scale factor normalized by the present epoch), occur deep in the radiation-dominated era, they act as a source term in the Einstein equation. Then, curvature perturbations experience logarithmic growth even on superhorizon scales. However, such anisotropic stress perturbations are compensated by neutrino anisotropic stress perturbations subsequent to neutrino decoupling, and therefore the enhancement of curvature perturbations stops. The resultant comoving curvature perturbations on superhorizon scales are evaluated as (\citet{Kojima:2009gw, Shaw:2009nf})
\begin{eqnarray}
{\cal R}_V({\bf k}) \approx 
R_\gamma \ln\left(\frac{\tau_\nu}{\tau_V}\right) 
\frac{3}{2} Q^j_{~i}(\hat{\bf k})
\Pi^i_{V j}({\bf k})~, \label{eq:RA}
\end{eqnarray}
where $\tau_\nu$ and $\tau_V$ are the conformal time of neutrino decoupling and the time of generation of the anisotropic stress perturbations, $R_\gamma \approx 0.6$ is the ratio of the energy density between photons and all relativistic particles, $Q^j_{~i}(\hat{\bf k}) \equiv - \hat{k^j}\hat{k_i} + \frac{1}{3} \delta^{j}_{~i}$, and $~\hat{}~$ denotes a unit vector.
\footnote{In the comoving gauge, ${\cal R}$ is related to the scalar metric perturbation as $g_{ij} = a^2 e^{2 {\cal R}} \delta_{ij}$.} The anisotropic stress perturbations normalized by the photon energy density, $\Pi^i_{V j}$, are defined as the traceless part of the energy momentum tensor: 
\begin{eqnarray}
T^i_{V j}({\bf k},\tau) \equiv \frac{\rho_{\gamma,0}}{a^4}
\left[\Delta_V({\bf k}) \delta^i_{~j} + \Pi^i_{V j}({\bf k})\right]~,
\end{eqnarray}
with $\rho_{\gamma,0}$ and $\Delta_V$ being the present photon energy density and the isotropic stress, respectively. The solution (\ref{eq:RA}) corresponds to the so-called passive mode in the context of the magnetized cosmology (\citet{Shaw:2009nf}). 

Equation (\ref{eq:RA}) indicates that the statistical property of ${\cal R}_V$ directly reflects that of $\Pi^i_{V j}$. $\Pi^i_{V j}$ consists of the square of the Gaussian vector fields as $\Pi^i_{V j}({\bf x}) = f_{V} V^i({\bf x}) V_j({\bf x})$, whose Fourier components are 
\begin{eqnarray}
\Pi^i_{V j}({\bf k}) = f_V \int \frac{d^3 {\bf k'}}{(2 \pi)^3} V^i({\bf k'}) V_j({\bf k} - {\bf k'}) ~, \label{eq:pi_def}
\end{eqnarray} 
and hence ${\cal R}_V$ obeys the non-Gaussian statistics and finite higher-order correlation functions can be produced. Here, the dimensionless coefficient $f_V$ behaves like the nonlinearity parameter of the local-type non-Gaussianity, $f_{\rm NL}$ (\citet{Komatsu:2001rj}). In Section~\ref{sec:EM}, by considering the electromagnetic action in the absence of the conformal invariance, we will see that $f_V$ corresponds to the running coupling of the electromagnetic action. 

\subsection{Bispectrum of curvature perturbations} 

If the power spectrum of $V_i$ is given by 
\begin{eqnarray}
\braket{V^i ({\bf k}) V_j ({\bf k'})} = (2 \pi)^3
 \frac{P_V(k)}{2} P^i_{~j}(\hat{\bf k}) \delta({\bf k} + {\bf k'}) ~,
\label{eq:power} 
\end{eqnarray}
with $P_V(k) \equiv \frac{2 \pi^2}{k^3} A_V (\frac{k}{k_*})^{n_V - 1}$ and $P^i_{~j}(\hat{\bf k}) \equiv \delta^i_{~j} - \hat{k}^i \hat{k}_j$, the bispectrum of ${\cal R}_V$ is expressed as 
\begin{eqnarray}
&&\Braket{\prod_{n=1}^3 {\cal R}_V({\bf k_n})} \nonumber \\ 
&& = \left[ \prod_{n=1}^3 R_\gamma \ln\left(\frac{\tau_\nu}{\tau_V}\right) 
\frac{3}{2} Q^{j_n}_{~i_n}(\hat{\bf k_n}) 
f_V \int d^3 {\bf k_n'} {P}_V(k_n') 
\right] \nonumber \\ 
&&\quad\times 
\delta({\bf k_1} - {\bf k_1'} + {\bf k_3'}) 
\delta({\bf k_2} - {\bf k_2'} + {\bf k_1'}) 
\delta({\bf k_3} - {\bf k_3'} + {\bf k_2'}) \nonumber \\
&&\quad\times P^{i_1}_{~j_2}(\hat{\bf k_1'}) P^{i_3}_{~j_1}(\hat{\bf k_3'}) P^{i_2}_{~ j_3}(\hat{\bf k_2'})~. \label{eq:ani_bis}
\end{eqnarray}
As can be seen, this involves an additional convolution due to the six-point function of the Gaussian vector fields. To obtain the actual bispectrum, we need to overcome this complicated convolution in a skillful manner. For the case that the power spectrum is nearly scale-invariant as $n_V \sim 1$, the contributions of the three poles at $k_1', k_2', k_3' \sim 0$ in the integrand dominate over those in equation (\ref{eq:ani_bis}). By picking up only these pole contributions in the same manner as in \citet{Shiraishi:2012rm}, we have 
\begin{eqnarray}
&&  \int d^3 {\bf k'} P_V(k') P^i_{~j}(\hat{\bf k'})  \nonumber \\ 
&& \to 
\beta_{n_V} \int_0^{k_*} k'^2 dk' P_V(k')  
\int d^2 \hat{\bf k'} 
P^i_{~j}(\hat{\bf k'})  \nonumber \\
&&= 2\pi^2 A_V \frac{\beta_{n_V} }{n_V - 1} 
\frac{8 \pi}{3}\delta^i_{~j} \ \ (n_V > 1)~, 
\end{eqnarray}
and, dealing with contractions, we obtain the reduced formula for $n_V \sim 1 (> 1)$ as 
\begin{eqnarray}
\Braket{\prod_{n=1}^3 {\cal R}_V({\bf k_n})} \equiv (2\pi)^3 B_{{\cal R}_V}(k_1, k_2, k_3) \delta\left(\sum_{n=1}^3 {\bf k_n}\right) ~,
\end{eqnarray}
where 
\begin{eqnarray}
B_{{\cal R}_V} 
&\approx& \left[\frac{3}{2} \pi F_V \right]^3 
 \frac{\beta_{n_V} }{n_V - 1} 
k_*^{2 (1 - n_V)} \frac{8\pi}{3}
S_V(k_1, k_2, k_3), \label{eq:curv_bis_V}  \\ 
F_V &\equiv& f_V A_V R_\gamma \ln\left(\frac{\tau_\nu}{\tau_V}\right) ~.
\end{eqnarray}
Here, we introduce $k_* = 0.002 {\rm Mpc}^{-1}$ and $\beta_{n_V}$ as a normalization scale used in the WMAP analysis (\citet{Komatsu:2010fb}) and a factor which should be determined by comparison with the exact bispectrum, respectively. From numerical studies, we obtain some specific values as $\beta_{1.1} = 0.785$ and $\beta_{1.01} = 1.036$.
\footnote{The relation between $\beta_{1.1}$ and $\alpha (= 0.335)$ of \citet{Shiraishi:2012rm} is given by 
\begin{eqnarray}
\beta_{1.1} = \left(\frac{10 {\rm Mpc}^{-1}}{k_*}\right)^{0.1} \alpha ~. \nonumber 
\end{eqnarray}
}

\begin{figure}
  \begin{center}
    \includegraphics[width = 8 cm]{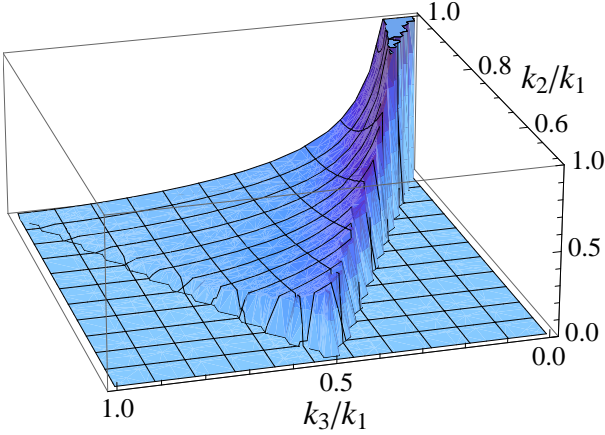}
  \end{center}
  \caption{The shape of $k_1^2 k_2^2 k_3^2 S_V (n_V = 1)$. Owing to the symmetric properties and triangle inequality, the plot range is restricted as $k_3 \leq k_2 \leq k_1$ and $|k_1 - k_2| \leq k_3 \leq k_1 + k_2$. It is shown that the bispectrum diverges in the squeezed limit as $k_3 \ll k_1 \approx k_2$.}
\label{fig:shape.eps}
\end{figure}

The shape function of the bispectrum is given by 
\begin{eqnarray}
S_V 
&=& \left[ \prod_{n=1}^3 Q^{j_n}_{~ i_n}(\hat{\bf k_n})  \right]
\nonumber \\
&& \times 
\left[ k_1^{n_V - 4} k_2^{n_V - 4} \delta^{i_1}_{~j_2} P^{i_3}_{~j_1}(\hat{\bf k_1}) P^{i_2}_{~j_3}(\hat{\bf k_2}) \right. \nonumber \\
&&\quad \left.
+ k_2^{n_V - 4} k_3^{n_V - 4} P^{i_1}_{~ j_2}(\hat{\bf k_2}) P^{i_3}_{~j_1}(\hat{\bf k_3}) \delta^{i_2}_{~j_3} \right. \nonumber \\
&&\quad \left. +  k_1^{n_V - 4} k_3^{n_V - 4} P^{i_1}_{~j_2}(\hat{\bf k_1}) \delta^{i_3}_{~j_1} P^{i_2}_{~j_3}(\hat{\bf k_3}) \right]
\nonumber \\ 
&=& \frac{1}{36} \left[ k_1^{n_V - 4 } k_2^{n_V - 4 } + {\rm 2 \ permutations}\right] \nonumber \\ 
&& - \frac{7}{216} \left[ k_1^{n_V-2} k_2^{n_V-6} + {\rm 5 \ permutations}\right] \nonumber \\ 
&&+ \frac{5}{216} \left[ k_1^{n_V-6} k_2^{n_V-4} k_3^2 + {\rm 5 \ permutations}\right] \nonumber \\
&& + \frac{1}{72} \left[ k_1^{-2} k_2^{n_V} k_3^{n_V-6} + {\rm 5 \ permutations}\right] \nonumber \\ 
&& - \frac{1}{72} \left[ k_1^{-2} k_2^{n_V-2} k_3^{n_V-4} + {\rm 5 \ permutations}\right] \nonumber \\
&& - \frac{1}{216} \left[ k_1^{n_V-6} k_2^{n_V-6} k_3^4 + {\rm 2 \ permutations}\right]~.  \label{eq:S_V}
\end{eqnarray}
This overall shape is illustrated in Fig.~\ref{fig:shape.eps}. We can see that the bispectrum for $n_V \sim 1$ is enhanced in the squeezed limit ($k_3 \ll k_1 \approx k_2$) and close to the local-type configuration. This may be due to the fact that the anisotropic stress perturbations $\Pi^i_{V j}$ are localized in real space like the local-type non-Gaussianity. The simplest description of curvature perturbations that gives exact local-type non-Gaussianity is conventionally written as ${\cal R}_{\rm loc}({\bf x}) = {\cal R}_{\rm g}({\bf x}) + \frac{3}{5}f_{\rm NL}[{\cal R}_{\rm g}^2({\bf x}) - \Braket{{\cal R}_{\rm g}^2({\bf x})}]$, with $f_{\rm NL}$ and ${\cal R}_{\rm g}$ denoting the local-type nonlinearity parameter and the Gaussian perturbation, respectively (e.g., \citet{Komatsu:2001rj}). Then, the local-type bispectrum is given by 
\begin{eqnarray}
\Braket{\prod_{n=1}^3 {\cal R}_{\rm loc}({\bf k_n})} \equiv (2\pi)^3 B_{{\cal R}_{\rm loc}}(k_1, k_2, k_3) \delta\left(\sum_{n=1}^3 {\bf k_n}\right) ~, 
\end{eqnarray}
with 
\begin{eqnarray} 
B_{{\cal R}_{\rm loc}} 
&=& \frac{6}{5} f_{\rm NL} 
\left[ P_{{\cal R}_g}(k_1) P_{{\cal R}_g}(k_2) + {\rm 2 \ permutations} \right] \nonumber \\  
&=& \frac{3}{5} f_{\rm NL} (2\pi^2 A_{\rm loc})^2 k_*^{2(1 - n_{\rm loc})} S_{\rm loc}(k_1, k_2, k_3) \label{eq:curv_bis_local}  ~, \\ 
S_{\rm loc} &\equiv& 2 k_1^{n_{\rm loc} - 4} k_2^{n_{\rm loc} - 4} + {\rm 2 \ permutations} ~. \label{eq:S_loc}
\end{eqnarray}
Here, $P_{{\cal R}_g}(k) \equiv \frac{2 \pi^2}{k^3} A_{\rm loc} (\frac{k}{k_*})^{n_{\rm loc} - 1}$ is the power spectrum of ${\cal R}_g$, $n_{\rm loc}$ is the spectral index, and $A_{\rm loc} = 2.43 \times 10^{-9}$ is a normalization factor obtained from the observational data of the scalar power spectrum (\citet{Komatsu:2010fb}). To quantify the resemblance between the shape functions $S$ and $S'$, the correlation coefficient is calculated, which is defined by $r(S, S') \equiv \frac{S \cdot S'}{\sqrt{(S \cdot S)(S' \cdot S')}}$ with (\cite{Babich:2004gb})
\begin{equation}
S \cdot S' = \int_0^1 dx_2 \int_{1-x_2}^1 d x_3 (x_2 x_3)^4 S(1, x_2, x_3) S'(1, x_2, x_3) ~.
\end{equation}
Here, $r$ reaches unity if $S$ closely resembles $S'$. Substituting equations (\ref{eq:S_V}) and (\ref{eq:S_loc}) into this equation, we obtain $r(S_V, S_{\rm loc}) = 0.96$ for $n_V = n_{\rm loc} \sim 1$, which indicates the shape similarity between $S_V$ and $S_{\rm loc}$.
\footnote{On the other hand, $S_V$ is nothing like the shape functions of other types of non-Gaussianity such as equilateral and orthogonal ones, because $r(S_V, S_{\rm eq}) = 0.03$ and $r(S_V, S_{\rm orth}) = -0.05$.} Actually, we can see that in the squeezed limit ($k \ll k'$), these shape functions have the same $k$ dependence as
\begin{eqnarray}
S_V(k, k', k') &\approx& \frac{2}{27} k^{n_V-4} k'^{n_V-4} 
 ~, \label{eq:S_V_sq} \\  
S_{\rm loc}(k, k', k')  
&\approx& 4 k^{n_{\rm loc}-4} k'^{n_{\rm loc}-4} 
 ~. \label{eq:S_loc_sq}
\end{eqnarray}
By using an approximate proportional relation, $S_V \approx \frac{S_V \cdot S_{\rm loc}}{S_{\rm loc} \cdot S_{\rm loc}}  S_{\rm loc} = 0.0238 S_{\rm loc}$, we can derive an approximate relation as 
\begin{eqnarray}
F_V^3 
&=& \frac{4}{15} \frac{n_V -1}{\beta_{n_V}} 
\frac{S_{\rm loc}}{S_V} f_{\rm NL} A_{\rm loc}^2  \nonumber \\
&\approx& f_{\rm NL}  
\times 
\begin{cases}
8.43 \times 10^{-18} & (n_V = 1.1) \\ 
6.39 \times 10^{-19} & (n_V = 1.01)
\end{cases}  ~. \label{eq:consist_rel}
\end{eqnarray} 
In the next section, we will determine whether or not this relation is valid even in the scale-dependent bias. 

\subsection{Interpretation as electromagnetic fields}\label{sec:EM}


Such non-Gaussian anisotropic stress perturbations can be created in the framework of inflationary models involving a conformally variant coupling between the scalar field $\phi$ and the vector field $A_\mu$, whose action is expressed as $S \supset - \int d^{4}x \frac{1}{4}\sqrt{-g} g^{\mu \lambda}g^{\nu \sigma}W(\phi)F_{\mu \nu}F_{\lambda \sigma}$ (e.g., \citet{Caldwell:2011ra, Motta:2012rn, Barnaby:2012tk}). Here, $F_{\mu \nu} \equiv \partial_\mu A_\nu - \partial_\nu A_\mu$, and $W(\phi)$ denotes the running coupling, which induces violation of the conformal invariance. Then, with the natural assumption that the inflaton falls into a stable state, $W(\phi)$ freezes out at the end of inflation as $W(\phi) \to W_{\rm I}$, and the conformal invariance is restored, we may have residual anisotropic stress perturbations generated from the electromagnetic fields (${\bf E}$ and ${\bf B}$) at later times. According to \citet{Shiraishi:2012xt}, if we take $f_V = -W_{\rm I}$, $V_i$ in equation~(\ref{eq:pi_def}) can be then equated to $E_i / \sqrt{4\pi \rho_{\gamma,0}}$ or $B_i / \sqrt{4\pi \rho_{\gamma,0}}$. 

From here, let us focus on the case that electric fields disappear owing to the Coulomb screening in the primordial plasma, and only magnetic fields survive at late times with $W_{\rm I} = 1$. This setting has often been considered in the context of cosmological magnetogenesis, but there remain outstanding issues such as the so-called strong coupling or backreaction problem (e.g. \citet{Demozzi:2009fu, Barnaby:2012tk, Fujita:2012rb}). In such a case, a conventional parametrization of the magnetic fields is given as (\citet{Shaw:2009nf, Shiraishi:2012rm})
\begin{eqnarray}
\braket{B^i ({\bf k}) B_j ({\bf k'})} &=& (2 \pi)^3
 \frac{P_B(k)}{2} P^i_{~j}(\hat{\bf k}) \delta({\bf k} + {\bf k'}) ~, \\ 
P_B(k) &\equiv& {A}_B k^{n_B} ~, \\
{A}_B &=& 
\frac{(2 \pi)^{n_B + 5} B^2_r}{\Gamma
 \left(\frac{n_B+3}{2}\right) k_r^{n_B + 3}} \ \ (n_B > -3)~, 
\end{eqnarray}
where $B_r$ denotes the magnetic field strength smoothed on the scale $r$, $k_r \equiv 2\pi / r$, $n_B$ is the spectral index of the magnetic power spectrum, and $\Gamma(x)$ is the Gamma function. Diverse analyses of cosmological phenomena such as the CMB anisotropy and the large-scale structure suggest constraints on these parameters as $B_{\rm 1Mpc} < {\cal O}(0.1 - 1){\rm nG}$ and $n_B \sim -3$ (e.g., \citet{Shaw:2010ea, Paoletti:2010rx, Shiraishi:2012rm, Yamazaki:2012pg, Pandey:2012ss, Paoletti:2012bb}). Accordingly, we hold the correspondence
\begin{eqnarray}
A_V &=& \frac{A_B k_*^{n_V-1}}{8 \pi^3 \rho_{\gamma, 0}}  ~, \\
n_V &=& n_B+4 ~.
\end{eqnarray}
Interestingly, for the magnetic case, because $f_V = -1$ and $B_{\rm 1Mpc} \geq 0$, the resultant bispectrum of curvature perturbations can mimic the local-type bispectrum only for $f_{\rm NL} \leq 0$. Thus, an approximate relation for the vector fields (\ref{eq:consist_rel}) is translated into  
\begin{eqnarray}
&& \left( \frac{B_{\rm 1Mpc}}{1 {\rm nG}} \right)
\left( \frac{\ln(\tau_\nu / \tau_B)}{\ln 10^{17}} \right)^{1/2} \nonumber \\ 
&&\approx \left(- f_{\rm NL}\right)^{1/6} 
\times
\begin{cases}
2.92 & (n_B = -2.9) \\
4.60 & (n_B = -2.99)
\end{cases}
~. \label{eq:consist_rel_mag}
\end{eqnarray}
This analytic evaluation implies that the bispectrum from the magnetic fields and the local-type bispectrum with negative $f_{\rm NL}$ have similar impacts also on the scale-dependent bias parameters, as shown in the following section. 

\section{Scale-dependent bias originating from vector fields}\label{sec:bias}

Recently, \citet{Matsubara:2012nc}, by applying iPT (\citet{Matsubara:2011ck}), constructed a more general formalism for the scale-dependent bias. In this section we employ this formalism to estimate the bias parameter from vector-induced non-Gaussian curvature perturbations given by $R_V$. 

\subsection{Formulation}


The bias parameters are often defined in the relation between the power spectrum of biased objects X, $P_{\rm X}$, and the linear matter power spectrum, $P_{\rm L}$:
\begin{eqnarray}
P_{\rm X}(k) \equiv 
\left[ b + \Delta b(k,M) \right]^2 P_{\rm L}(k)~, \label{eq:bias_def}
\end{eqnarray}
 where $b$ denotes the scale-independent Eulerian linear bias parameter and $\Delta b$ is the scale-dependent bias parameter arising from the primordial non-Gaussianity. In the literature, there is another notation for the bias parameters, which links the cross-correlation between the linearized matter density field and objects $X$, $P_{\rm L X}$, to $P_{\rm L}$ as 
\begin{eqnarray}
P_{\rm L X}(k) \equiv 
\left[b + \Delta b(k,M)\right] P_{\rm L}(k)~. \label{eq:bias_def_cross}
\end{eqnarray}
These two $\Delta b$s are identical under the situation that the non-Gaussianity is so weak that the contribution of the trispectrum is negligible (\citet{Matsubara:2012nc}). In other words, considering a highly non-Gaussian source, $\Delta b$ in equation~(\ref{eq:bias_def}) is affected by the higher-order correlations and may differ from $\Delta b$ in equation~(\ref{eq:bias_def_cross}) on large scales (\citet{Yokoyama:2011qr, Yokoyama:2012az}). Here, we shall blink this fact tentatively and formulate the scale-dependent bias arising from the bispectrum of curvature perturbations.


According to \citet{Matsubara:2012nc}, the scale-dependent bias parameter is expressed as
\begin{eqnarray}
\Delta b (k, M) \approx \frac{\sigma_M^2}{2 \delta_c^2} 
\left[ A_2(M) {\cal I}(k, M) + A_1(M) \frac{\partial {\cal I}(k, M)}{\partial \ln \sigma_M} \right], \label{eq:bias}
\end{eqnarray}
where $\delta_c = 1.686$ is the critical overdensity, and 
\begin{eqnarray}
{\cal I}(k, M) &=& 
\frac{1}{\sigma_M^2 P_{\rm L}(k)}  
\int \frac{d^3 {\bf k'}}{(2 \pi)^3} W(k' R) W(|{\bf k} - {\bf k'}| R) \nonumber \\ 
&&\times B_{\rm L}(k, k', |{\bf k} - {\bf k'}|) 
~. 
\end{eqnarray}
The coefficients $A_1$ and $A_2$ are expanded by the Lagrangian bias parameters $b_1^{\rm L}$ and $b_2^{\rm L}$ as 
\begin{eqnarray}
A_1(M) &=& 1 + \delta_c b_1^{\rm L}(M) ~, \\ 
A_2(M) &=& 2 + 2 \delta_c b_1^{\rm L}(M) + \delta_c^2 b_2^{\rm L}(M) ~.
\end{eqnarray}
Following the notation of the halo model, we have introduced the Lagrangian radius $R$ satisfying 
$M = \frac{4\pi}{3} \Omega_{m0} \rho_{c0} R^3$ 
, which is equivalent to 
\begin{eqnarray}
\frac{R}{\rm Mpc} = \left[ \frac{M}{1.162 \times 10^{12} h^2 M_\odot \Omega_{m0}} \right]^{1/3} ~.
\end{eqnarray}
Here, $\rho_{c0}$ and $\Omega_{m0}$ are the present values of the critical density and the matter density parameter, respectively, $h \equiv H_0 / (100 {\rm km/sec/Mpc}) $ with $H_0$ being the Hubble constant, and $M_\odot = 1.989 \times 10^{30} {\rm kg}$ is the mass of the sun. Then, the density variance is defined by  
\begin{eqnarray}
\sigma_M^2 = \int \frac{k^2 d k}{2 \pi^2} W^2(kR) P_{\rm L}(k) ~, 
\end{eqnarray}
where we choose a top-hat window function as
\begin{eqnarray}
W(x) = \frac{3 j_1(x)}{x} ~,
\end{eqnarray}
with $j_\ell(x)$ being the spherical Bessel function. 

The Lagrangian bias parameters, $b_1^{\rm L}$ and $b_2^{\rm L}$, depend on the shape of the mass function. In our numerical calculation, we adopt the fitting formulae derived from the ``MICE mass function'', which has been estimated from the data of MICE simulations (\citet{Crocce:2009mg}), as 
\begin{eqnarray}
b_1^{\rm L}(M) &=& \frac{1}{\delta_c} 
\left( \frac{2c_3}{\sigma_M^2} - \frac{c_1}{1 + c_2 \sigma_M^{c_1}} \right) ~, \\ 
b_2^{\rm L}(M) &=& \frac{1}{\delta_c^2} 
\left[ \frac{4c_3^2}{\sigma_M^4} - \frac{2c_3}{\sigma_M^2} 
- \frac{c_1 (4c_3 / \sigma_M^2 - c_1 + 1)}{1 + c_2 \sigma_M^{c_1}} \right] ~, 
\end{eqnarray}
with $c_1 = 1.37 a^{0.15}$, $c_2 = 0.3 a^{0.084}$ and $c_3 = 1.036 a^{0.024}$. 

The linear matter power spectrum, $P_{\rm L}$, and the bispectrum sourced from the primordial non-Gaussianity, $B_{\rm L}$, are expressed as 
\begin{eqnarray}
P_{\rm L}(k) &=& {\cal M}_{\cal R}^2(k) P_{\cal R}(k)~, \\
B_{\rm L}(k_1, k_2, k_3) &=& \left[ \prod_{n=1}^3 {\cal M}_{\cal R}(k_n) \right] B_{\cal R}(k_1, k_2, k_3)
 ~, 
\end{eqnarray}
where ${\cal M}_{\cal R}$ denotes the conversion function involving the information of the linear evolution of the matter contrast and is parametrized as 
\begin{eqnarray}
{\cal M}_{\cal R}(k) = \frac{2}{5} D(a) 
\frac{k^2 T(k)}{H_0^2 \Omega_{m0}} ~.
\end{eqnarray}
The matter transfer function $T(k)$ and the growth factor $D(a)$ are realized by such fitting functions as (\citet{Lahav:1991wc, Weinberg:2008zzc})
\begin{eqnarray}
&& T(k) = \frac{\ln[1 + (0.124 \kappa)^2]}{(0.124 \kappa)^2} \nonumber \\ 
&&\qquad \times 
\sqrt{\frac{ 1 + (1.257 \kappa)^2 + (0.4452 \kappa)^4 + (0.2197 \kappa)^6 }
{ 1 + (1.606 \kappa)^2 + (0.8568 \kappa)^4 + (0.3927 \kappa)^6} } 
~,
\end{eqnarray}
with 
\begin{eqnarray}
\kappa &=& \frac{k \sqrt{\Omega_{r0}}}{H_0 \Omega_{m0}} 
\left[ \alpha + \frac{1 - \alpha}{1 + (0.43 k s)^4} \right]^{-1} ~, \\ 
\alpha &=& 1 - 0.328 \ln (431 \Omega_{m0} h^2) \frac{\Omega_{b0}}{\Omega_{m0}} \nonumber \\ 
&& + 0.38 \ln(22.3 \Omega_{m0} h^2) \left( \frac{\Omega_{b0}}{\Omega_{m0}} \right)^2 ~, \\ 
s &=& \frac{44.5 \ln [9.83 / (\Omega_{m0} h^2)]}{ \sqrt{1 + 10(\Omega_{b0} h^2)^{3/4}} } {\rm Mpc} ~, 
\end{eqnarray}
and 
\begin{eqnarray}
D(a) &=&  \frac{5}{2} a \Omega_m \nonumber \\ 
&&\times 
\left[ \Omega_m^{4/7} - \Omega_\Lambda 
+ \left( 1 + \frac{\Omega_m}{2} \right) 
 \left( 1 + \frac{\Omega_\Lambda}{70} \right) 
\right]^{-1}.
\end{eqnarray}
Here, $\Omega_{\Lambda 0}, \Omega_{b 0}$ and $\Omega_{r 0}$ are the present density parameters of the cosmological constant, baryons and radiation, respectively, $\Omega_m(a) \equiv \frac{\Omega_{m0} }{\Omega_{m0} + a^3 \Omega_{\Lambda 0}}$, and $\Omega_\Lambda(a) \equiv \frac{a^3 \Omega_{\Lambda 0}}{\Omega_{m0} + a^3 \Omega_{\Lambda 0}}$. In the matter dominated era, $D(a)$ converges to $a$.  

To obtain the scale-dependent bias from the primordial vector fields or the local-type non-Gaussianity, we only input several power spectra and bispectra described in the previous section to equation~(\ref{eq:bias}). 

\subsection{Analysis}\label{sec:result}

Let us focus on the numerical analysis of the scale-dependent bias from the vector fields, $\Delta b^V$. For comparison, we also calculate the scale-dependent bias from the local-type bispectrum, $\Delta b^{\rm loc}$. 
In what follows, the value of the power spectrum of curvature perturbations is derived from observations by the {\it WMAP} experiment, namely (\citet{Komatsu:2010fb})
\begin{eqnarray}
P_{{\cal R}}(k) \equiv \frac{2 \pi^2}{k^3} A_{\cal R}
\left(\frac{k}{k_*}\right)^{n_{\cal R} - 1} ~, 
\end{eqnarray}
with $A_{\cal R} = A_{\rm loc} = 2.43 \times 10^{-9}$ and $n_{\cal R} = 0.963$. In our case, the vector fields create not only the bispectrum but also the power spectrum of curvature perturbations, namely $P_{{\cal R}_V}$, as calculated in Appendix~\ref{appen:power_vec}. In this sense, $P_{{\cal R}_V}$ becomes part of $P_{\cal R}$ along with the power spectra from other sources such as the cross-bispectra between metric perturbations and the electromagnetic fields (e.g. \citet{Shiraishi:2012xt, Kunze:2012fd}). Because of the lack of an explicit dependence of $P_{\cal R}$ on $F_V$, $\Delta b^V$ is simply proportional to $F_V^3$. 


\begin{figure}
  \begin{center}
    \includegraphics[width = 8 cm]{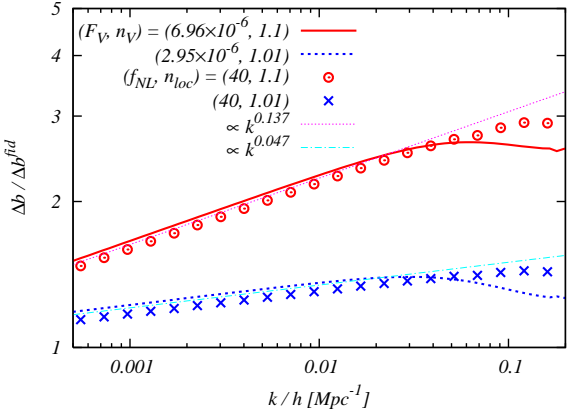}
  \end{center}
  \caption{Scale-dependent bias parameters normalized by $\Delta b^{\rm loc}$ for $(f_{\rm NL}, n_{\rm loc}) = (40, 0.963)$ (corresponding to $\Delta b^{\rm fid}$) at redshift $z \equiv a^{-1} - 1 = 1$. Red solid and blue dotted lines correspond to $\Delta b^V / \Delta b^{\rm fid}$, and red circles and blue crosses denote $\Delta b^{\rm loc} / \Delta b^{\rm fid}$ with $f_{\rm NL} = 40$ when $n_V = n_{\rm loc} = 1.1, 1.01$, respectively. The parameters of these amplitudes are taken to satisfy the approximate relation (\ref{eq:consist_rel}) as seen in the graph legends. The other related parameters are fixed as $\Omega_{m 0} = 0.275$, $\Omega_{\Lambda 0} = 1 - \Omega_{m 0}$, $h = 0.7$ and $M/M_{\odot} = 10^{14}$. Magenta fine dotted and cyan chain lines with $k$ dependence as in the graph legends are plotted to check of the scaling relations (\ref{eq:delb_scale}).}
\label{fig:ratio}
\end{figure}

In Fig.~\ref{fig:ratio}, we plot the ratios of $\Delta b^V$ and $\Delta b^{\rm loc}$ to $\Delta b^{\rm fid} \equiv \Delta b^{\rm loc}(f_{\rm NL} = 40, n_{\rm loc} = n_{\cal R})$ for the parameters satisfying the approximate relation (\ref{eq:consist_rel}) when $f_{\rm NL} = 40$ and $n_V = n_{\rm loc} = 1.1, 1.01$. From this figure, it can be seen that the red solid (blue dotted) line is in agreement with the red circles (blue crosses) on large scales. This fact indicates that the approximate relation (\ref{eq:consist_rel}) is true even in the scale-dependent bias. On the other hand, the deviation on small scales may arise from the tiny difference of $k$ dependence between the primordial bispectra (\ref{eq:S_V}) and (\ref{eq:S_loc}). This may become a key signal for extracting information on the vector fields from the contamination of other local-type non-Gaussian sources.


From the red solid and blue dotted curves of this figure, it can be seen that a difference of tilt between $\Delta b^V(n_V = 1.1)$ and $\Delta b^V(n_V = 1.01)$ is evident. This different $k$ dependence can be analytically estimated as follows. From the $k$-dependent part in equation~(\ref{eq:bias}), we have
\begin{eqnarray}
\Delta b(k) &\propto& \frac{1}{P_{\rm L}(k)} 
\int \frac{d^3 {\bf k'}}{(2 \pi)^3} W(k' R) W(|{\bf k} - {\bf k'}| R) \nonumber \\ 
&&\times B_{\rm L}(k, k', |{\bf k} - {\bf k'}|) ~.
\end{eqnarray} 
In the large-scale limit, the integrand of this equation reduces to the form under the squeezed limit ($k \ll k'$); that is, 
\begin{eqnarray}
W(|{\bf k} - {\bf k'}| R) &\approx& W(k' R) ~, \\ 
B_{\rm L}(k, k', |{\bf k} - {\bf k'}|) 
&\approx& {\cal M}_{\cal R}(k) {\cal M}_{\cal R}^2 (k') B_{\cal R}(k, k', k') ~. 
\end{eqnarray}
Using the squeezed-limit forms of the shape functions (\ref{eq:S_V_sq}) and (\ref{eq:S_loc_sq}), the large-scale approximation of the matter transfer function as $T(k) \approx 1$, namely, ${\cal M}_{\cal R}(k) \propto k^2$, and $P_{\cal R}(k) \propto k^{n_{\cal R} - 4}$, we obtain the scaling relation of each bias on large scales as  
\begin{eqnarray}
\begin{split}
\Delta b^V(k) &\propto k^{n_V - n_{\cal R} - 2} \\ 
\Delta b^{\rm loc}(k) &\propto k^{n_{\rm loc} - n_{\cal R} - 2} ~, 
\\ 
\Delta b^{\rm fid}(k) &\propto k^{- 2} ~. \label{eq:delb_scale}
\end{split}
\end{eqnarray}  
The $k$ dependence derived from these equations, $\Delta b / \Delta b^{\rm fid} \propto k^{0.137}$ for $n_V = n_{\rm loc} = 1.1$ and $k^{0.047}$ for $n_V = n_{\rm loc} = 1.01$, fits well to the curves and points of Fig.~\ref{fig:ratio} on large scales. 


\begin{figure}
  \begin{center}
    \includegraphics[width = 8 cm]{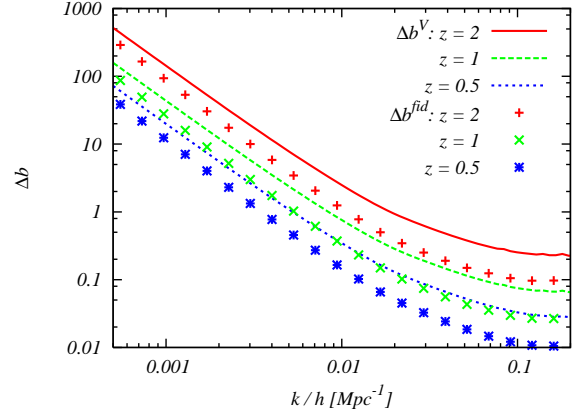}
  \end{center}
  \caption{Scale-dependent bias parameters for $z = 2, 1, 0.5$. 
The three lines correspond to $\Delta b^V$ with $(F_V, n_V) = (6.96 \times 10^{-6}, 1.1)$, and the three types of symbols correspond to $\Delta b^{\rm fid}$, for the three values of $z$, respectively (see the graph legends). The related parameters are identical to the values represented in Fig.~\ref{fig:ratio}.}
\label{fig:delb_per_z}
\end{figure}

Figure~\ref{fig:delb_per_z} shows the redshift dependence of $\Delta b^{V}$ and $\Delta b^{\rm fid}$. We can see that as $z$ evolves, the overall amplitudes of $\Delta b^V$ and $\Delta b^{\rm fid}$ simply decrease while their shapes are maintained. Such multi-$z$ information will become more valuable when we analyse the tomographic data from {\it BOSS} and forthcoming {\it EUCLID} experiments (\citet{Ross:2012sx, Laureijs:2011mu}).


\begin{figure}
  \begin{center}
    \includegraphics[width = 8 cm]{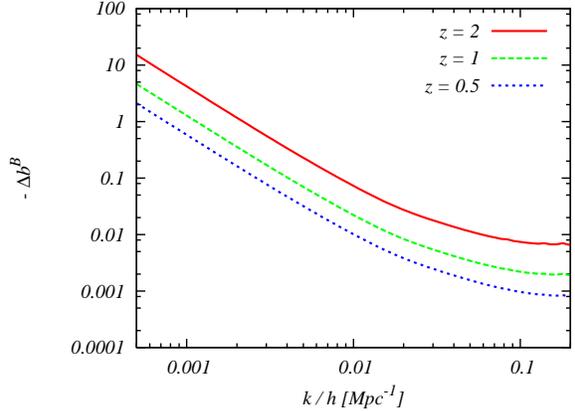}
  \end{center}
  \caption{Scale-dependent bias parameter from the primordial magnetic fields for $z = 2, 1, 0.5$ (the vertical axis describes $- \Delta b^B$). Here, the parameters of the magnetic fields are taken to be consistent with observations as $(B_{\rm 1Mpc},\tau_\nu / \tau_B, n_B) = (3 {\rm nG}, 10^{17}, -2.9)$. The other parameters are identical to the values represented in Fig.~\ref{fig:ratio}.}
\label{fig:delb_per_z_B}
\end{figure}

In the remainder of the paper, we examine how the magnetic fields affect the scale-dependent bias. As shown in Section.~\ref{sec:EM}, the bispectrum of curvature perturbations induced by magnetic anisotropic stress perturbations corresponds to the bispectrum only for negative $f_{\rm NL}$. This dependence appears also in the scale-dependent bias. Fig.~\ref{fig:delb_per_z_B} describes the scale-dependent bias from the magnetic fields,  $\Delta b^B$, for each redshift. The magnetic parameters are chosen as $(B_{\rm 1Mpc},\tau_\nu / \tau_B, n_B) = (3 {\rm nG}, 10^{17}, -2.9)$, consistent with the current observational bounds (e.g., \citet{Shaw:2010ea, Paoletti:2010rx, Shiraishi:2012rm, Yamazaki:2012pg, Paoletti:2012bb, Pandey:2012ss}). In this case, $\Delta b^B$ is equivalent to $\Delta b^V$ for $(F_V, n_V) = (- 2.15 \times 10^{-6}, 1.1)$. Translating these magnetic fields into the local-type non-Gaussianity for $n_{\rm loc} = 1.1$ using equation~(\ref{eq:consist_rel_mag}), we have $f_{\rm NL} = -1.2$. From this figure, it can be confirmed that $\Delta b^B$ has negative values at all $k$. This negative impact with the deviation from the fiducial $k$ dependence as $\Delta b^{\rm fid} \propto k^{-2}$ can change more or less the entire shape of the scale-dependent bias parameter and become a key feature for probing the primordial magnetic fields. 
If the vector fields are the primordial magnetic fields, one may be concerned about additional influences from small-scale non-Gaussian matter fluctuations driven by the Lorentz force, which is the so-called magnetic-compensated mode (\cite{Shaw:2009nf}). The scale-dependent bias becomes significant when some amount of fluctuations on large scales are correlated with small scale ones through the bispectrum in the squeezed limit. Fluctuations from the magnetic-compensated mode, however, are suppressed on large scales and therefore we do not expect the fluctuations to affect the clustering of haloes on large scales.

\subsection{On higher-order effects}

In the above evaluation, we took into account only the effects of the bispectrum of curvature perturbations. Concern remains, however, about the contribution of the higher-order correlation because curvature perturbations driven by the vector fields (\ref{eq:RA}) are Gaussian-squared fields, that is, highly non-Gaussian fields. 
In \citet{Yokoyama:2011qr}, the halo-halo bias was computed when curvature perturbations originated in the Gaussian-squared scalar fields. For this case, we have a consistency relation between the nonlinearity parameters associated with the bispectrum and trispectrum of curvature perturbations as 
$\tau_{\rm NL} \sim 500 f_{\rm NL}^{4/3}$. If the angular dependence due to the vector fields is negligible, this relation may be applicable to our case. Then, considerable trispectrum contribution may modify the halo-halo bias from the vector fields on large scales, as pointed out in \citet{Yokoyama:2011qr, Baumann:2012bc, Yokoyama:2012az}. However, this effect is sensitive to the coefficient of $f_{\rm NL}^{4/3}$ in the above relation, and hence detailed calculations with the complicated angular dependence are required for precise discussion. It is an issue for future work. 

\section{Summary and discussion}

In this paper we have investigated the scale-dependent bias that arises from the non-Gaussianity of curvature perturbations induced by Gaussian vector fields. This non-Gaussianity induces the local-type bispectrum of curvature perturbations. We found an approximate relation between the strength of the anisotropic stress perturbations from the vector fields $F_V$ and the local-type nonlinearity parameter $f_{\rm NL}$, as shown in equation~(\ref{eq:consist_rel}). Through numerical calculations, we demonstrated that, as this relation predicts, the scale-dependent bias from the vector field is in agreement with that from the local-type non-Gaussianity on large scales, while we observed deviation from the local-type bias on small scales. This feature may be useful for seeking evidence of the primordial vector fields from observations of the scale-dependent bias. Furthermore, the $k$ dependence of the scale-dependent bias directly reflects the spectral tilt of the power spectrum of the vector fields as shown in equation~(\ref{eq:delb_scale}). Thus, through the $k$ dependence, we will also be able to approach the shape of the primordial vector field. By interpreting the primordial vector fields as the primordial magnetic fields, we determined that the scale-dependent bias has negative values. This is an interesting consequence and helpful for constraining the primordial magnetic fields. Comparing the theoretical results involving the contribution of the higher-order correlations with the observational data leads to a greater understanding of the nature of the vector fields. 

\section*{Acknowledgments}

This work was supported in part by a Grant-in-Aid for JSPS Research under grant nos 22-7477 (MS) and 24-2775 (SY), a Grant-in-Aid for Scientific Research under grant nos. 24340048 (KI) and 24540267 (TM), and a Grant-in-Aid for Nagoya University Global COE Program 'Quest for Fundamental Principles in the Universe: from Particles to the Solar System and the Cosmos', from the Ministry of Education, Culture, Sports, Science and Technology of Japan. We also acknowledge the Kobayashi-Maskawa Institute for the Origin of Particles and the Universe, Nagoya University for providing computing resources that were useful in conducting the research reported in this paper.

\bibliography{paper}

\begin{thebibliography}{}

\bibitem[\protect\citeauthoryear{Ackerman, Carroll \& Wise}{Ackerman
  et~al.}{2007}]{Ackerman:2007nb}
Ackerman L.,  Carroll S.~M.,    Wise M.~B.,  2007, Phys. Rev., D75, 083502,
  \eprint{astro-ph/0701357}

\bibitem[\protect\citeauthoryear{Babich, Creminelli \& Zaldarriaga}{Babich
  et~al.}{2004}]{Babich:2004gb}
Babich D.,  Creminelli P.,    Zaldarriaga M.,  2004, JCAP, 0408, 009,
  \eprint{astro-ph/0405356}

\bibitem[\protect\citeauthoryear{Bamba \& Sasaki}{Bamba \&
  Sasaki}{2007}]{Bamba:2006ga}
Bamba K.,  Sasaki M.,  2007, JCAP, 0702, 030, \eprint{astro-ph/0611701}

\bibitem[\protect\citeauthoryear{Bamba \& Yokoyama}{Bamba \&
  Yokoyama}{2004}]{Bamba:2003av}
Bamba K.,  Yokoyama J.,  2004, Phys.Rev., D69, 043507,
  \eprint{astro-ph/0310824}

\bibitem[\protect\citeauthoryear{Barnaby, Moxon, Namba, Peloso, Shiu
  et~al.,}{Barnaby et~al.}{2012}]{Barnaby:2012xt}
Barnaby N.,  Moxon J.,  Namba R.,  Peloso M.,  Shiu G.,    et~al., 2012,
  Phys.Rev., D86, 103508, \eprint{1206.6117}

\bibitem[\protect\citeauthoryear{Barnaby, Namba \& Peloso}{Barnaby
  et~al.}{2011}]{Barnaby:2011vw}
Barnaby N.,  Namba R.,    Peloso M.,  2011, JCAP, 1104, 009, \eprint{1102.4333}

\bibitem[\protect\citeauthoryear{Barnaby, Namba \& Peloso}{Barnaby
  et~al.}{2012}]{Barnaby:2012tk}
Barnaby N.,  Namba R.,    Peloso M.,  2012, Phys.Rev., D85, 123523,
  \eprint{1202.1469}

\bibitem[\protect\citeauthoryear{Barnaby \& Peloso}{Barnaby \&
  Peloso}{2011}]{Barnaby:2010vf}
Barnaby N.,  Peloso M.,  2011, Phys.Rev.Lett., 106, 181301, \eprint{1011.1500}

\bibitem[\protect\citeauthoryear{Bartolo, Komatsu, Matarrese \& Riotto}{Bartolo
  et~al.}{2004}]{Bartolo:2004if}
Bartolo N.,  Komatsu E.,  Matarrese S.,    Riotto A.,  2004, Phys. Rept., 402,
  103, \eprint{astro-ph/0406398}

\bibitem[\protect\citeauthoryear{Bartolo, Matarrese, Peloso \&
  Ricciardone}{Bartolo et~al.}{2013}]{Bartolo:2012sd}
Bartolo N.,  Matarrese S.,  Peloso M.,    Ricciardone A.,  2013, Phys.Rev.,
  D87, 023504, \eprint{1210.3257}

\bibitem[\protect\citeauthoryear{Baumann, Ferraro, Green \& Smith}{Baumann
  et~al.}{2012}]{Baumann:2012bc}
Baumann D.,  Ferraro S.,  Green D.,    Smith K.~M.,  2012, \eprint{1209.2173}

\bibitem[\protect\citeauthoryear{Becker \& Huterer}{Becker \&
  Huterer}{2012}]{Becker:2012je}
Becker A.,  Huterer D.,  2012, Phys.Rev.Lett., 109, 121302, \eprint{1207.5788}

\bibitem[\protect\citeauthoryear{Becker, Huterer \& Kadota}{Becker
  et~al.}{2012}]{Becker:2012yr}
Becker A.,  Huterer D.,    Kadota K.,  2012, JCAP, 1212, 034,
  \eprint{1206.6165}

\bibitem[\protect\citeauthoryear{Brown \& Crittenden}{Brown \&
  Crittenden}{2005}]{Brown:2005kr}
Brown I.,  Crittenden R.,  2005, Phys. Rev., D72, 063002,
  \eprint{astro-ph/0506570}

\bibitem[\protect\citeauthoryear{Caldwell, Motta \& Kamionkowski}{Caldwell
  et~al.}{2011}]{Caldwell:2011ra}
Caldwell R.~R.,  Motta L.,    Kamionkowski M.,  2011, Phys.Rev., D84, 123525,
  \eprint{1109.4415}

\bibitem[\protect\citeauthoryear{Crocce, Fosalba, Castander \&
  Gaztanaga}{Crocce et~al.}{2010}]{Crocce:2009mg}
Crocce M.,  Fosalba P.,  Castander F.~J.,    Gaztanaga E.,  2010,
  Mon.Not.Roy.Astron.Soc., 403, 1353, \eprint{0907.0019}

\bibitem[\protect\citeauthoryear{Demozzi, Mukhanov \& Rubinstein}{Demozzi
  et~al.}{2009}]{Demozzi:2009fu}
Demozzi V.,  Mukhanov V.,    Rubinstein H.,  2009, JCAP, 0908, 025,
  \eprint{0907.1030}

\bibitem[\protect\citeauthoryear{Demozzi \& Ringeval}{Demozzi \&
  Ringeval}{2012}]{Demozzi:2012wh}
Demozzi V.,  Ringeval C.,  2012, JCAP, 1205, 009, \eprint{1202.3022}

\bibitem[\protect\citeauthoryear{Dimastrogiovanni, Bartolo, Matarrese \&
  Riotto}{Dimastrogiovanni et~al.}{2010}]{Dimastrogiovanni:2010sm}
Dimastrogiovanni E.,  Bartolo N.,  Matarrese S.,    Riotto A.,  2010,
  Adv.Astron., 2010, 752670, \eprint{1001.4049}

\bibitem[\protect\citeauthoryear{Fujita \& Mukohyama}{Fujita \&
  Mukohyama}{2012}]{Fujita:2012rb}
Fujita T.,  Mukohyama S.,  2012, JCAP, 1210, 034, \eprint{1205.5031}

\bibitem[\protect\citeauthoryear{Hikage \& Matsubara}{Hikage \&
  Matsubara}{2012}]{Hikage:2012bs}
Hikage C.,  Matsubara T.,  2012, Mon.Not.Roy.Astron.Soc., 425, 2187,
  \eprint{1207.1183}

\bibitem[\protect\citeauthoryear{Jain \& Sloth}{Jain \&
  Sloth}{2012}]{Jain:2012ga}
Jain R.~K.,  Sloth M.~S.,  2012, Phys.Rev., D86, 123528, \eprint{1207.4187}

\bibitem[\protect\citeauthoryear{Kanno, Soda \& Watanabe}{Kanno
  et~al.}{2009}]{Kanno:2009ei}
Kanno S.,  Soda J.,    Watanabe M.-a.,  2009, JCAP, 0912, 009,
  \eprint{0908.3509}

\bibitem[\protect\citeauthoryear{Karciauskas, Dimopoulos \& Lyth}{Karciauskas
  et~al.}{2009}]{Karciauskas:2008bc}
Karciauskas M.,  Dimopoulos K.,    Lyth D.~H.,  2009, Phys.Rev., D80, 023509,
  \eprint{0812.0264}

\bibitem[\protect\citeauthoryear{Kojima, Kajino \& Mathews}{Kojima
  et~al.}{2010}]{Kojima:2009gw}
Kojima K.,  Kajino T.,    Mathews G.~J.,  2010, JCAP, 1002, 018,
  \eprint{0910.1976}

\bibitem[\protect\citeauthoryear{Komatsu}{Komatsu}{2010}]{Komatsu:2010hc}
Komatsu E.,  2010, Class. Quant. Grav., 27, 124010, \eprint{1003.6097}

\bibitem[\protect\citeauthoryear{Komatsu et~al.,}{Komatsu
  et~al.}{2011}]{Komatsu:2010fb}
Komatsu E.,  et~al., 2011, ApJS, 192, 18, \eprint{1001.4538}

\bibitem[\protect\citeauthoryear{Komatsu \& Spergel}{Komatsu \&
  Spergel}{2001}]{Komatsu:2001rj}
Komatsu E.,  Spergel D.~N.,  2001, Phys. Rev., D63, 063002,
  \eprint{astro-ph/0005036}

\bibitem[\protect\citeauthoryear{Kunze}{Kunze}{2013}]{Kunze:2012fd}
Kunze K.~E.,  2013, JCAP, 1302, 009, \eprint{1209.4570}

\bibitem[\protect\citeauthoryear{Lahav, Lilje, Primack \& Rees}{Lahav
  et~al.}{1991}]{Lahav:1991wc}
Lahav O.,  Lilje P.~B.,  Primack J.~R.,    Rees M.~J.,  1991,
  Mon.Not.Roy.Astron.Soc., 251, 128

\bibitem[\protect\citeauthoryear{Laureijs, Amiaux, Arduini, Augueres,
  Brinchmann et~al.,}{Laureijs et~al.}{2011}]{Laureijs:2011mu}
Laureijs R.,  Amiaux J.,  Arduini S.,  Augueres J.-L.,  Brinchmann J.,
  et~al., 2011, \eprint{1110.3193}

\bibitem[\protect\citeauthoryear{Martin \& Yokoyama}{Martin \&
  Yokoyama}{2008}]{Martin:2007ue}
Martin J.,  Yokoyama J.,  2008, JCAP, 0801, 025, \eprint{0711.4307}

\bibitem[\protect\citeauthoryear{Matsubara}{Matsubara}{2011}]{Matsubara:2011ck}
Matsubara T.,  2011, Phys.Rev., D83, 083518, \eprint{1102.4619}

\bibitem[\protect\citeauthoryear{Matsubara}{Matsubara}{2012}]{Matsubara:2012nc}
Matsubara T.,  2012, Phys.Rev., D86, 063518, \eprint{1206.0562}

\bibitem[\protect\citeauthoryear{Motta \& Caldwell}{Motta \&
  Caldwell}{2012}]{Motta:2012rn}
Motta L.,  Caldwell R.~R.,  2012, Phys.Rev., D85, 103532, \eprint{1203.1033}

\bibitem[\protect\citeauthoryear{Pandey \& Sethi}{Pandey \&
  Sethi}{2013}]{Pandey:2012ss}
Pandey K.~L.,  Sethi S.~K.,  2013, Astrophys.J., 762, 15, \eprint{1210.3298}

\bibitem[\protect\citeauthoryear{Paoletti \& Finelli}{Paoletti \&
  Finelli}{2011}]{Paoletti:2010rx}
Paoletti D.,  Finelli F.,  2011, Phys.Rev., D83, 123533, \eprint{1005.0148}

\bibitem[\protect\citeauthoryear{Paoletti \& Finelli}{Paoletti \&
  Finelli}{2012}]{Paoletti:2012bb}
Paoletti D.,  Finelli F.,  2012, \eprint{1208.2625}

\bibitem[\protect\citeauthoryear{Ross, Percival, Carnero, Zhao, Manera
  et~al.,}{Ross et~al.}{2012}]{Ross:2012sx}
Ross A.~J.,  Percival W.~J.,  Carnero A.,  Zhao G.-b.,  Manera M.,    et~al.,
  2012, \eprint{1208.1491}

\bibitem[\protect\citeauthoryear{Sefusatti, Liguori, Yadav, Jackson \&
  Pajer}{Sefusatti et~al.}{2009}]{Sefusatti:2009xu}
Sefusatti E.,  Liguori M.,  Yadav A.~P.,  Jackson M.~G.,    Pajer E.,  2009,
  JCAP, 0912, 022, \eprint{0906.0232}

\bibitem[\protect\citeauthoryear{Shandera, Dalal \& Huterer}{Shandera
  et~al.}{2011}]{Shandera:2010ei}
Shandera S.,  Dalal N.,    Huterer D.,  2011, JCAP, 1103, 017,
  \eprint{1010.3722}

\bibitem[\protect\citeauthoryear{Shaw \& Lewis}{Shaw \&
  Lewis}{2010}]{Shaw:2009nf}
Shaw J.~R.,  Lewis A.,  2010, Phys.Rev., D81, 043517, \eprint{0911.2714}

\bibitem[\protect\citeauthoryear{Shaw \& Lewis}{Shaw \&
  Lewis}{2012}]{Shaw:2010ea}
Shaw J.~R.,  Lewis A.,  2012, Phys.Rev., D86, 043510, \eprint{1006.4242}

\bibitem[\protect\citeauthoryear{Shiraishi}{Shiraishi}{2012}]{Shiraishi:2012sn}
Shiraishi M.,  2012, JCAP, 1206, 015, \eprint{1202.2847}

\bibitem[\protect\citeauthoryear{Shiraishi, Nitta, Yokoyama \&
  Ichiki}{Shiraishi et~al.}{2012}]{Shiraishi:2012rm}
Shiraishi M.,  Nitta D.,  Yokoyama S.,    Ichiki K.,  2012, JCAP, 1203, 041,
  \eprint{1201.0376}

\bibitem[\protect\citeauthoryear{Shiraishi, Saga \& Yokoyama}{Shiraishi
  et~al.}{2012}]{Shiraishi:2012xt}
Shiraishi M.,  Saga S.,    Yokoyama S.,  2012, JCAP, 1211, 046,
  \eprint{1209.3384}

\bibitem[\protect\citeauthoryear{Slosar, Hirata, Seljak, Ho \&
  Padmanabhan}{Slosar et~al.}{2008}]{Slosar:2008hx}
Slosar A.,  Hirata C.,  Seljak U.,  Ho S.,    Padmanabhan N.,  2008, JCAP,
  0808, 031, \eprint{0805.3580}

\bibitem[\protect\citeauthoryear{Smidt, Amblard, Byrnes, Cooray, Heavens
  et~al.,}{Smidt et~al.}{2010}]{Smidt:2010ra}
Smidt J.,  Amblard A.,  Byrnes C.~T.,  Cooray A.,  Heavens A.,    et~al., 2010,
  Phys.Rev., D81, 123007, \eprint{1004.1409}

\bibitem[\protect\citeauthoryear{Sorbo}{Sorbo}{2011}]{Sorbo:2011rz}
Sorbo L.,  2011, JCAP, 1106, 003, \eprint{1101.1525}

\bibitem[\protect\citeauthoryear{Suyama \& Yokoyama}{Suyama \&
  Yokoyama}{2012}]{Suyama:2012wh}
Suyama T.,  Yokoyama J.,  2012, Phys.Rev., D86, 023512, \eprint{1204.3976}

\bibitem[\protect\citeauthoryear{Valenzuela-Toledo \&
  Rodriguez}{Valenzuela-Toledo \& Rodriguez}{2010}]{ValenzuelaToledo:2009nq}
Valenzuela-Toledo C.~A.,  Rodriguez Y.,  2010, Phys.Lett., B685, 120,
  \eprint{0910.4208}

\bibitem[\protect\citeauthoryear{Verde}{Verde}{2010}]{Verde:2010wp}
Verde L.,  2010, Adv.Astron., 2010, 768675, \eprint{1001.5217}

\bibitem[\protect\citeauthoryear{Watanabe, Kanno \& Soda}{Watanabe
  et~al.}{2010}]{Watanabe:2010fh}
Watanabe M.-a.,  Kanno S.,    Soda J.,  2010, Prog.Theor.Phys., 123, 1041,
  \eprint{1003.0056}

\bibitem[\protect\citeauthoryear{Watanabe, Kanno \& Soda}{Watanabe
  et~al.}{2011}]{Watanabe:2010bu}
Watanabe M.-a.,  Kanno S.,    Soda J.,  2011, Mon.Not.Roy.Astron.Soc., 412,
  L83, \eprint{1011.3604}

\bibitem[\protect\citeauthoryear{{Weinberg}}{{Weinberg}}{2008}]{Weinberg:2008z%
zc}
{Weinberg} S.,  2008, {Cosmology}.
Oxford University Press,
  \adsurl{http://adsabs.harvard.edu/abs/2008cosm.book.....W}

\bibitem[\protect\citeauthoryear{Widrow}{Widrow}{2002}]{Widrow:2002ud}
Widrow L.~M.,  2002, Rev. Mod. Phys., 74, 775, \eprint{astro-ph/0207240}

\bibitem[\protect\citeauthoryear{Yamazaki, Kajino, Mathew \& Ichiki}{Yamazaki
  et~al.}{2012}]{Yamazaki:2012pg}
Yamazaki D.~G.,  Kajino T.,  Mathew G.~J.,    Ichiki K.,  2012, Phys.Rept.,
  517, 141, \eprint{1204.3669}

\bibitem[\protect\citeauthoryear{Yokoyama}{Yokoyama}{2011}]{Yokoyama:2011qr}
Yokoyama S.,  2011, JCAP, 1111, 001, \eprint{1108.5569}

\bibitem[\protect\citeauthoryear{Yokoyama \& Matsubara}{Yokoyama \&
  Matsubara}{2013}]{Yokoyama:2012az}
Yokoyama S.,  Matsubara T.,  2013, Phys.Rev., D87, 023525, \eprint{1210.2495}

\bibitem[\protect\citeauthoryear{Yokoyama \& Soda}{Yokoyama \&
  Soda}{2008}]{Yokoyama:2008xw}
Yokoyama S.,  Soda J.,  2008, JCAP, 0808, 005, \eprint{0805.4265}

\end{thebibliography}

\appendix

\section{Power spectrum of curvature perturbations induced from vector fields}\label{appen:power_vec}

\begin{figure}
  \begin{center}
    \includegraphics[width = 8 cm]{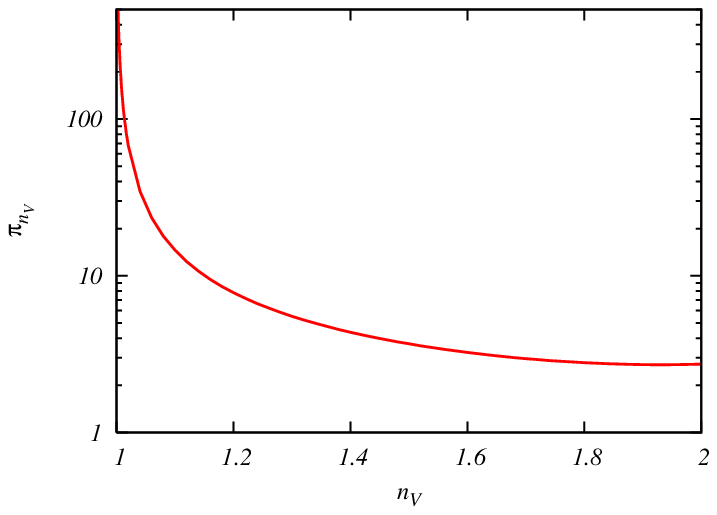}
  \end{center}
  \caption{Dependence of $\pi_{n_V}$ on $n_V$.}
\label{fig:pi_V.eps}
\end{figure}

Here we present the analytic formula for the power spectrum of curvature perturbations arising from the anisotropic stress perturbations composed of the square of Gaussian vector fields. Notations in this section are based on \cite{Shaw:2009nf}. From the equations in Section~\ref{sec:non-gaussianity}, the power spectrum is formed as
\begin{eqnarray} 
\Braket{\prod_{n=1}^2 {\cal R}_V({\bf k_n})} &=& (2\pi)^3 
P_{{\cal R}_V}(k_1) \delta\left( \sum_{n=1}^2 {\bf k_n} \right) ~, \\
P_{{\cal R}_V}(k) 
&=& \frac{2 \pi^2}{k^3} F_V^2 \frac{\pi_{n_V}}{4} 
\left( \frac{k}{k_*} \right)^{2( n_V-1)} ~.
\end{eqnarray} 
A factor $\pi_{n_V}$ arises from the convolution in terms of $P_V(k)$ as
\begin{eqnarray}
\pi_{n_V} &=& \frac{k^{-2n_V + 5}}{2\pi} 
\int d^3 {\bf k'} k'^{n_V - 4} |{\bf k} - {\bf k'}|^{n_V - 4} \nonumber \\
&&\times \left[ 1 - \frac{3}{4} (\beta^2 + \gamma^2) 
+ \frac{9}{4} \beta^2 \gamma^2 
- \frac{3}{2} \mu \gamma \beta + \frac{1}{4} \mu^2
\right]  \nonumber \\ 
&=& 
\int_0^\infty du u^{n_V-2} 
\int_{-1}^1 d\gamma (1 - 2 u \gamma + u^2)^{(n_V-4)/2} \nonumber \\ 
&&\times \left[ 1 - \frac{3}{4} (\beta^2 + \gamma^2) 
+ \frac{9}{4} \beta^2 \gamma^2 
- \frac{3}{2} \mu \gamma \beta + \frac{1}{4} \mu^2
\right]~, \nonumber \\
\end{eqnarray}
where the parameters correspond to 
\begin{eqnarray}
u &=& \frac{k'}{k} ~, \\   
\gamma &=& \hat{\bf k} \cdot \hat{\bf k'} ~, \\ 
\mu &=& \hat{\bf k'} \cdot \widehat{{\bf k} - {\bf k'}} 
= \frac{\gamma - u}{\sqrt{1 + u^2 - 2 u \gamma}}~, \\ 
\beta &=& \hat{\bf k} \cdot \widehat{{\bf k} - {\bf k'}} 
= \frac{1 - u \gamma}{\sqrt{1 + u^2 - 2 u \gamma}} ~, \\ 
|{\bf k} - {\bf k'}|^2 &=& k^2 
\left( 1 + u^2 - 2 u \gamma \right) ~.
\end{eqnarray}
While the integral over $\gamma$ can be analytically evaluated, we have to rely on numerical calculation in the integral over $u$. In Fig.~\ref{fig:pi_V.eps}, we show $\pi_{n_V}$ for $1 < n_V < 2$. It can be seen that as $n_V$ decreases, the contribution around the pole dominates and $\pi_{n_V}$ is boosted. 

\end{document}